\documentstyle[twocolumn,aps,epsf,subfigure]{revtex}

\begin{document}
\twocolumn[\hsize\textwidth\columnwidth\hsize\csname@twocolumnfalse\endcsname
\title{
Location-Sensitive Measurement of the Local Fluctuation of Driven Vortex Density in Bi$_2$Sr$_2$CaCu$_2$O$_y$
}
\author{
A. Maeda
}
\address{
Department of Basic Science, The University of Tokyo,\\
3-8-1, Komaba, Meguro-ku, Tokyo 153-8902, Japan \\
CREST, Japan Science and Technology Corporation (JST),\\
4-1-8, Honcho, Kawaguchi, 332-0012, Japan
}
\author{
T. Tsuboi, R. Abiru, Y. Togawa, H. Kitano, K. Iwaya
}
\address{
Department of Basic Science, The University of Tokyo,\\
3-8-1, Komaba, Meguro-ku, Tokyo 153-8902, Japan
}
\author{
T. Hanaguri
}
\address{
Department of Advanced Materials Science, The University of Tokyo,\\
7-3-1, Hongo, Bunkyo-ku, Tokyo 113-8656, Japan
}

\date{\today}
\maketitle
\newpage
\begin{abstract}

To investigate the dynamics of driven vortices in superconductors, noise in the local vortex density was investigated in the mixed state of a high-$T_c$ superconductor, Bi$_2$Sr$_2$CaCu$_2$O$_y$, using a two-dimensional electron gas (2DEG) micro-Hall probe array.  We studied the cross-correlation function, together with the auto-correlation function, both parallel and perpendicular to the direction of flow of the vortices.
The broadband noise (BBN) did not have large spatial correlations.  This suggests that the BBN is due to the fluctuation of the local vortex density generated by bulk pinning centers under the area of each probe.
On the other hand, the narrow-band noise (NBN) (with the the time scale of the transit time of vortices) had large translational correlations.
These definitely shows that the NBN was generated by semimacroscopic imperfections like the surface, and macroscopic line defects {\it etc.}.
In relation to the dynamic phase diagram, large BBN was observed when the vortices started moving.  The dependence of the spatial correlations on the direction of the array relative to the direction of the driving current suggested that plastic flow was present when the large BBN was observed.
The gross features of our data agree well with some of the theoretically proposed dynamical phase diagrams of vortices in superconductors.

\end{abstract}

\vspace{1em}
]

\newpage
\narrowtext

\section{INTRODUCTION}

Since the discovery of high-temperature superconductivity in cuprates, the physics of the vortex matter has attracted greater attention than before the high-$T_c$ era.
This is because, in high-$T_c$ superconductors, various important energy scales such as thermal energy, pinning energy, elastic energy, and interaction between vortices are comparable with each other, which leads to a rich variety of new physics in the mixed state of high-$T_c$ superconductors\cite{Blatter}.
In particular, the physics of current-driven vortices has attracted much recent attention\cite{George}.
This subject concerns the physics of a driven system under random pinning, and much work has been focused on the nature of the moving state and also on the dynamic phase diagram, namely the phase diagram in the $H-T-F$ ($F$ is the driving force) space.
The physics of driven vortices has many common aspects to the dynamics of charge- and spin-density waves (CDW and SDW) in quasi-one dimensional materials, and to the dynamics of Wigner crystals in correlated materials, and also to solid friction.
This means that the subject probes a quite general, important issue in condensed-matter physics.

In high-$T_c$ superconductors, a complex equilibrium phase diagram, particularly the presence of the first-order transition (FOT) of the vortex lattice\cite{George,FOTBSC,FOTYBC}, makes the problem more complicated and challenging.

In a textbook of superconductivity\cite{Tinkham}, the situation has been considered where a vortex lattice moves elastically under a driving force.  If there is finite pinning, flux creep takes place.  On the other hand, when there is almost no pinning, flux flow occurs.
In both cases, vortices were considered to move in an elastic bundle.
Recently, however, a more complex form of vortex motion has been considered.
Several theoretical studies\cite{Vinokur,Balents,Doussal,Balents2} and numerical simulations\cite{Jensen,Zimanyi,Dominiguez,Ryu,Olson} have discussed in detail the possible dynamical phases of driven vortices in superconductors.
Many of them agree with each other regarding the following aspects.
If the driving force increases, depinned vortices start moving.  When the driving force is small, vortices move in a plastic manner (plastic flow (PF)), where there are channels in which vortices move with a finite velocity, whereas in other channels vortices remain pinned.  Thus, between moving channels and static channels, there are dislocations in the vortex lattice.
With further increasing driving current, vortices tend to re-order.
This is a common feature of the theoretical studies listed above.
However, regarding the exact detail, there has been no consensus on what the dynamical phase of the vortices should be.

Koshelev and Vinokur\cite{Vinokur} proposed a dynamic melting under a finite driving force.  Through this dynamic melting, the stationary vortex lattice changes into the moving vortex lattice.  In addition, the critical field (or the critical temperature) corresponding to the FOT of the vortex lattice in the equilibrium state did not depend on the driving force.

Giamarchi and Le Doussal\cite{Doussal} proposed another ordered phase, the moving-Bragg-glass.  This phase is characterized by the presence of translational order whose spatial correlation decays in a power law both parallel and perpendicular to the translational motion of the vortices.
According to this theory, the transition between the Bragg glass and the moving-Bragg-glass is not a phase transition, but rather a gradual crossover.
They also argued that the Bragg glass changes into the moving Bragg glass directly, without a region of plastic flow, if the disorder is weak.
On the other hand, if the disorder is strong, the equilibrium phase is the vortex glass.
After going through the PF, it changes into the moving transverse glass, which does not posses translational order.

Phenomenologically, the moving-Bragg-glass phase can be characterized by the presence of the so-called washboard oscillation,
which is a periodic velocity modulation caused by the interaction between an almost uniform translation of a periodic object and pinning centers.  Thus, the frequency, $f$, of the washboard oscillation is simply given as $f$=$v/a$, where $v$ is the velocity and $a$ is the spatial periodicity of the object.  This phenomenon is well known in CDW and SDW systems.\cite{Gruner}.

On the other hand, Balents {\it et al.}\cite{Balents} proposed a smectic phase which lacks translational order in the longitudinal direction of the flow even in the case of weak disorder.  This was based on the experimental fact that the washboard noise had not been observed in the vortices of superconductors before, except in a system with artificially introduced strong periodic pinning centers\cite{Mart} and in the thermally-excited motion of a limited number of vortices\cite{Troy}.

Quite recently, washboard oscillation was observed in the noise spectrum generated by driven vortex motion under random pinning in the superconductor Bi$_2$Sr$_2$CaCu$_2$O$_y$\cite{Togawa}.  This supports the notion of the development of translational order along the flow direction, and should force reconsideration of current theories. 
However, a detailed picture of the vortex motion at each point of the dynamic phase diagram still remains a serious open question.

To elucidate this issue, several experimental efforts have been made\cite{Troy,Shobo1,Henderson,Yaron,Pardo}.  In particular,
various imaging techniques\cite{Troy,Yaron,Pardo}
have established the existence of various states of driven vortices in conventional superconductors.
However, for high-$T_c$ superconductors, almost nothing has been clarified, since
these probes are applicable only over a limited parameter range (temperature, magnetic field, and driving force)\cite{Fendrich,Tsuboi1,TMatsuda}.
For high-$T_c$ superconductors, we chose a slightly different approach; local-density-noise (LDN) measurement\cite{Tsuboi2,Tsuboishort,Maeda}.
Observation of low-frequency-noise has been a popular tool to study the vortex motion both in conventional and high-$T_c$ superconductors\cite{Clem,Bat1,Bat2,Danna,Safar}, and such noise has been considered to have some relationship with various low-frequency phenomena\cite{Paltier,Gordiev}.
This conduction noise represents {\it the velocity fluctuation} averaged over a macroscopic sample.
However, what we discuss below is {\it the local-density fluctuation} of vortices.
This probe has a merit that we can discuss how the local fluctuations generated at different places are correlated with each other.
  Yeh {\it et al.}\cite{WJYeh} already studied the density fluctuation using a SQUID technique both in a conventional superconductor and in a high-$T_c$ superconductor.  However, they did not concentrate on the dependence on the location nor on the spatial correlations.

In this paper, we present the results of a study of the LDN of vortices generated in a small area in the sample, measured by a micro-Hall probe array, with a special emphasis on the spatial correlations of the noise generated at different places both parallel and perpendicular to the flow direction.
Using this technique, we tried to clarify (1) the nature of the vortex state at each point of the dynamical phase space (2) the nature of the re-ordered phase (smectic phase {\it vs} moving-Bragg-glass phase) (3) characteristic aspects for high-$T_c$ cuprate superconductors.

\section{EXPERIMENTAL}

\subsection{Sample preparation and Characterization}

The samples were single crystals of Bi$_2$Sr$_2$CaCu$_2$O$_y$ (Bi-2212) made by the floating-zone method.
Almost optimally doped crystals were obtained by appropriate post annealing.
They were cut into rectangular pieces with typical dimensions of 1.5$\times$0.5 mm$^2$ in the ${ab}$ plane and 0.02 mm
thickness along the $c$-axis.  Four probe dc resistivity measurements showed no indication of other phases such as the so-called 2223 phase (Bi$_2$Sr$_2$Cu$_2$Cu$_3$O$_y$).  The resistivity just above the critical temperature $T_c$ was $\sim$400$\mu\Omega$cm. The critical temperature, $T_c$, ranged between 82 K and 89 K, depending on the crystals, which is probably due to very slight differences in the chemical composition of the material.
We do not believe that these differences of $T_c$ between different pieces of crystal should affect the essential physics discussed below.
Table~\ref{samples} lists the samples which were used in the measurements described in this paper.

\subsection{Local Density Noise Measurement}

\subsubsection{Definition of the quantities we measured}

The central significance of this paper is the time ($t$) dependent noise (fluctuation) of the local vortex density at the site $i$, $x_i(t)$, which is correlated to its Fourier component, $X_i(f)$ ($f$ is frequency) as
\begin{equation}
X_i(f) = \int_{-\infty}^{\infty}x_i(t){\rm e}^{-j2\pi ft}{\rm d}t.
\end{equation}
The power spectral density, $S_{ii}(f)$, is the square average of $X_i(f)$.
\begin{equation}
S_{ii}(f) \equiv \lim_{T\to\infty}<\frac{X_i(f)X_i^*(f)}{T}>,
\end{equation}
where * represents the complex conjugate, and $<>$ represents the ensemble average.
According to the Wiener-Kintchine theorem, $S_{ii}(f)$ is related to the auto-correlation function, $A_{ii}(t)$,
\begin{equation}
A_{ii}(t) = \lim_{T\to\infty}\frac{1}{T} \int_{-T/2}^{T/2}x_i(t_0)x_i(t_0+t){\rm d}t_0,
\end{equation}
as
\begin{equation}
S_{ii}(f) = \int_{-\infty}^{\infty}A_{ii}(t){\rm e}^{-j2\pi ft}{\rm d}t.
\end{equation}

On the other hand, the cross-correlation function between the density noise at the $i$ site, $x_i(t)$, and that at the $j$-site, $x_j(t)$, $A_{ij}(t)$ is defined as
\begin{equation}
A_{ij}(t) = \lim_{T\to\infty}\frac{1}{T}\int_{-T/2}^{T/2}x_i(t_0)x_j(t_0+t){\rm d}t_0,
\end{equation}
and represents how the fluctuations generated at these two different places are correlated spatially.
The cross-spectral density
\begin{equation}
S_{ij}(f) \equiv \lim_{T\to\infty}<\frac{X_i(f)X_j^*(f)}{T}>
\end{equation}
is found to be equal to the Fourier transform of $A_{ij}(t)$ as
\begin{equation}
S_{ij}(f) = \int_{-\infty}^{\infty}A_{ij}(t){\rm e}^{-j2\pi ft}{\rm d}t.
\end{equation}
Below, we use the normalized cross-spectral density,
\begin{equation}
H_{ij}(f) = \frac{S_{ij}(f)}{\sqrt{S_{ii}(f)S_{jj}(f)}}.
\end{equation}
Note that the cross-spectral density, $S_{ij}$, and the normalized cross-spectral density, $H_{ij}$, are complex quantities.  So, we can represent
\begin{equation}
H_{ij}(f) \equiv h_{ij}(f) {\rm e}^{i\theta(f)}.
\end{equation}
The amplitude, $h_{ij}$, is called {\it coherence}.  If $h_{ij}$ is unity, that means the fluctuations generated at the two places are completely correlated.  On the other hand, if $h_{ij}$ is zero, these fluctuations are uncorrelated.

\subsubsection{Measurements}

The LDN of the vortices was measured by a GaAs/GaAl$_{1-x}$As$_x$ 2DEG micro-Hall probe array (active area; 5 $\times$ 5 $\mu$m$^2$ or 15 $\times$ 15 $\mu$m$^2$, spacing between centers of each probe; 30 $\mu$m) placed on the flat surface of the specimen.
There are typically 4-6 active areas in one array.
The excitation level of the Hall probe is typically 10 $\mu$A$\sim$ 100 $\mu$A.
The voltage generated by each active area is proportional to the local vortex density below the area.  Thus, measurement of the average voltage tells us the local magnetization.
After a differential preamplification, the noise voltage of the Hall sensors was analyzed by an HP-35665 FFT analyzer.  Our measurement sensitivity was limited by the background noise level of the probe, which was typically 0.0005 G$/\sqrt{{\rm Hz}}$ above 200 Hz.

On the back side of the crystal, four electrodes were attached to measure the electrical resistivity.
Thus we can investigate both the LDN and electrical resistivity simultaneously in the same experimental run.
Our experimental setup is shown schematically in Fig.~\ref{expsetup}.

Spatial correlations of the noise were measured simply by putting the noise voltage generated at different places into two different signal inputs of the analyzer, and calculating the cross-spectral density, $S_{ij}(f)$.
The array could be aligned in the direction parallel or perpendicular to the driven current.  
By doing so, the spatial correlations of the density noise could be measured both parallel and perpendicular to the flow direction (Fig.~\ref{expprlvsperp}).

As for the data presentation, no background (noise sepctrum at zeor current density and zero magnetic field) subtraction was made in the power spectral density data, other than Figs.~\ref{BBNPSeach} and \ref{NBNPSeach}.
Cross correlations were calculated with using the data without backgrownd subtraction.

The noise measurements were performed around the FOT for various driving forces.
Since the driving force density {\bf f} is expressed as
\begin{equation}
{\bf f} = {\bf j} \times {\bf B},
\end{equation}
where {\bf j} and {\bf B} are the driving current density and magnetic field, respectively, we changed either {\bf j} or {\bf B} in the experiment.  For convenience, typically we swept magnetic field with a very slow rate under a fixed driving dc current.
It should be noted, however, in the {\bf B} swept experiment, the ``equilibrium" condition was also changed.
For experiments which changed the driving force under fixed ``equilibrium" parameters ($T$ and {\bf B}), a {\bf j}-swept experiment under constant {\bf B} is preferable.

In all measurements, the magnetic field was applied perpendicular to the CuO$_2$ plane.

\section{EXPERIMENTAL RESULTS}

\subsection{Total Features}

Figure~\ref{resist} shows the resistivity- and zero-current magnetization data of three typical samples as a function of magnetic field under a fixed current.
For instance, in Fig.~\ref{resist}~(a), a clear anomaly was observed at 43 Oe, which corresponds to the FOT in the vortex lattice.  Around the magnetization anomaly, the resistivity changed rapidly.  Although not shown in the figure, the magnetization anomaly was found to be independent of the driving current.  This behavior is in good agreement with our previous result\cite{Tsuboi1}, which is also consistent with theoretical predictions\cite{Vinokur}.  However, the field value where the resistivity changes very rapidly depends on the driving current.  Thus, as was already clarified in ref.~\cite{Tsuboi1}, the resistivity is not a good measure of the FOT for Bi-2212.

Figure~\ref{noisetotal} shows a typical LDN spectrum as a function of driving current under fixed magnetic field and temperature.  For this sample (\# 72c6), the magnetic field value of 31.8 Oe corresponds to the solid phase in the zero-current limit.
Extra noise appeared with increasing driving current, and the noise spectrum consisted of two kinds of noise, broadband noise (BBN) and narrow-band noise (NBN) which has a peak at some frequency and its harmonics.
Previously\cite{Tsuboi2}, we investigated the noise spectra as a function of magnetic field {\bf H} under fixed current density {\bf j}, and found that
the BBN took its maximum just before the resistivity onset.
In addition, the NBN appeared after the BBN intensity began to decrease.
On the other hand, the data shown in Fig.~\ref{noisetotal}, where the experiment was done for various current densities {\bf j}'s under a fixed magnetic field {\bf H}, have total features which appear different from what were seen in ref.\cite{Tsuboi2}.
In the data shown in Fig.~\ref{noisetotal}, (1) the BBN survived up to high driving current densities ({\it eg,} up to 240 - 288 A/cm$^2$) and (2) the NBN appeared even at a driving current density where the BBN was still large ({\it eg,} up to 144 -240 A/cm$^2$).
These differences are apparently strange, for both kinds of experiments are similar in the sense that the driving force was increased.
However, sweeping the magnetic field under constant current density crosses the magnetic field anomaly corresponding to the FOT in the zero-current limit.  On the other hand, sweeping the current density under constant magnetic field does not cross the magnetization anomaly.
This difference leads to the apparent differences between the data in ref.\cite{Tsuboi2} and those presented in Fig.~\ref{noisetotal}.
This will be shown more clearly in the next subsection.
Unlike this difference, one common feature was that the frequency of the NBN shifted to higher frequencies with increasing driving force in both experiments.

The data in Fig.~\ref{noisetotal} show that at higher current densities ({\it eg.}, 432 A/cm$^2$) another source of noise generation occured.
This probably corresponds to the presence of pinning centers with different pinning forces.

It should be also noted that the presence or the absence of the NBN depended on the sample. For example, among the three samples shown in Table~\ref{samples}, only two samples other than {\#}92r3 exhibited the NBN.
This suggests that the origin of the NBN is sensitive to the sample quality.  Alternatively, the NBN generation might be related to an extrinsic mechanism unconnected with the bulk pinning properties.

In the following subsection, we will discuss the BBN in more detail.

\subsection{Broadband Noise}

The inset of Fig.~\ref{BBNTdep} shows the equi-noise contour in the current density ({\bf j}) {\it vs} magnetic field ({\bf H}) plane measured at various temperatures.
The differences discussed above can be understood more easily in this figure.
That is, if {\bf H} is increased under constant {\bf j}, the noise power takes a maximum at a certain field value, and decreases rapidly at the field corresponding to the FOT in the zero-current limit.
On the other hand, if one increases {\bf j} under constant {\bf H}, the noise power also takes a maximum value.  However, it still remains at a rather large value.
For instance, if we increase {\bf j} along the line of {\bf H}=125 Oe at 44 K, the noise power takes a maximum at around 790 A/cm$^2$.  After this maximum, however, it remains large up to $\sim$1200 A/cm$^2$.
This corresponds to the apparent asymmetry already discussed between the data in Fig.~\ref{noisetotal} and those in ref.~\cite{Tsuboi2}.
We consider that the asymmetry in these two experiments is characteristic of high-$T_c$ cuprate superconductor, where the FOT exists in the equilibrium phase diagram.
Except for the presence of the FOT, we consider that the behavior of the noise power is basically symmetric in the {\bf j} - {\bf H} plane.

When the driving force density {\bf f}$=${\bf j}$\times${\bf B} is constant, {\bf j}$\propto$1/{\bf B}.
If we can regard that {\bf B} $\sim$ {\bf H}, then the constant driving force, {\bf f}, corresponds to the relationship, {\bf j}$\propto$1/{\bf H}.
  The equi-noise power data are found to be approximately along this {\it constant}-{\bf f} curve at each temperature.  Furthermore, the data suggest that the noise power takes its maximum on another {\it constant}-{\bf f} curve.

To compare the noise intensity with resistivity data, the equi-noise contour is shown together with the equi-velocity contour at 44 K (the main panel of Fig.~\ref{BBNTdep}), where the velocity $v$ was estimated as $v={\rho}j/B=E/B=V/LB$ ($E$ is electric field, V is voltage between the potential electrodes, $L$ is the distance between the potential electrodes,
$\rho$ is the resistivity, $j$ is the current density, 
and $B$ is the magnetic field).
In this figure, 7.5$\times$10 cm/s corresponds to 10$^{-9}$ $\Omega$cm, which is about the sensitivity limit of our resistivity measurement.
  The data show that the noise takes its maximum just before the resistivity onset (``velocity onset").
Although the data are not shown in Fig.~\ref{BBNTdep}, essentially the same results were obtained at all other temperatures investigated.
Thus, it was confirmed that large BBN was generated when the vortices start to move with a critical driving force probably corresponding to the critical current density, {\bf j}$_c$, at various points of the three dimensional phase diagram of $H$-$T$-$F$.

Figure~\ref{BBNIinvert} compares the noise spectra taken at driving current densities with the same magnitude but different directions.  Although the NBN position differed very slightly, the BBN spectra for different driving-force directions were found to be almost identical.

As was already discussed previously\cite{Tsuboi2}, the presence of the finger-print effect and the independence of the driving-force direction suggest that the BBN was generated by the bulk pinning centers located under each probe.
To see this more clearly, cross-correlation measurements should be powerful probes, which will be shown in section III D.

\subsection{Narrow-band Noise}

Although the BBN is suggested to have a bulk origin, the experimental features of the NBN support a rather different origin for the NBN.
Below, we will list the experimental features of the NBN.

(1) The presence or the absence of the NBN is dependent on the samples, which is in sharp contrast to the fact that the BBN was observed in all the samples investigated.

(2) The time scale of the NBN was found to be the transit time of vortices\cite{Tsuboi2}.
This is quite different from the time scale of the NBN observed in the conduction noise, which was the washboard noise\cite{Togawa}.

(3) As was shown just above, the frequency of the NBN depended on the driving-force direction.
A similar phenomenon was already reported in the conduction noise data of YBa$_2$Cu$_3$O$_y$, although the magnetic field was applied parallel to the CuO$_2$ plane\cite{Danna}.

All of these data suggest that the NBN generation is related to the presence of semi-macroscopic defect.
Again, to confirm this idea, cross-correlation measurements are effective.

\subsection{Spatial Correlation Measurement}

\subsubsection{BBN vs NBN}

We clarified that the BBN reached a maximum just before the resistivity onset at various points in the $H$-$T$-$F$ diagram.
This suggests that the BBN is related to the depinning process.
On the other hand, the plastic flow that was predicted close to the resistivity onset might also be the origin of a large BBN, since the opening and closing of the moving channels leads to the LDN of vortices.
To see which is more probable in the present case, spatial-correlation measurements of the noise are useful.
As was mentioned above, spatial correlation measurements of the noise are also important to determine the origins of the NBN.
If we choose a value for the current where the NBN is clearly observed in the spectrum, then the spatial correlations of the NBN can be studied.
On the other hand, if we choose the current value at a level where only the BBN is observed, then the spatial correlations of the BBN can be studied.

We will discuss the cross-correlation of the BBN first.
Figure~\ref{BBNPSeach} shows the power spectral density of the BBN of sample {\#}72c6 taken at 4 different sites.
Only the BBN was observed at all the sites at these values of {\bf H} and {\bf j}.
The cross-correlations were investigated between these 4
different sites.
Figure~\ref{crosstotal} shows the coherence $h_{ij}\equiv \frac{\vert S_{ij}\vert}{\sqrt{S_{ii}S_{jj}}}$ taken between two neighboring sites  when the sensor array was placed in the direction parallel to the flow direction, where $S_{ii}$ and $S_{jj}$ are the autocorrelation of the noise voltage at the sites i and j, respectively, and $S_{ij}$ is the cross-correlation of the noise voltage between the i site and the j site.
The data in Figs.~\ref{crosstotal}~(a) and (b) show that the cross-correlation of the BBN is $\sim$0.02 at most above 160 Hz.  Thus, the spatial correlation of the BBN was found to be rather small even between the neighboring sites.
It should be noted that the data taken between site 3 and site 4 (Fig.~\ref{crosstotal}~(c)) and the low-frequency part of  Fig.~\ref{crosstotal}~(b) show slightly larger spatial correlations.
By investigating the cross-correlations in various places in other samples, we found that the behavior seen between site 3 and the neighboring sites is exceptional.  We believe that the general behavior is that the BBN shows a small spatial correlation.

Figure~\ref{crossfar} shows the cross-correlation between sites which are farther from each other.
Although the data taken between site 1 and site 3 show slightly larger correlations only at very low frequencies, the overall result is that the BBN shows rather small spatial correlations also between these sites.

Next, let us move on to the NBN.
Figure~\ref{NBNPSeach} shows the power spectra of the NBN of sample {\#}72c6 taken at 4 different sites.
In all spectra, the NBN was clealy observed.
Figure~\ref{NBNcrossfar} shows the coherence $h_{ij}$ taken between two different sites  with different distances (30, 60 and 90 $\mu$m) when the sensors were put in the direction parallel to the flow direction.
It is remarkable that the NBN shows a very large spatial correlation even between the farthest sites (1-4), which is in sharp contrast to the BBN data.
This supports our previous conclusion that the NBN is related to the transfer of the density fluctuation generated at some point in the sample, definitely.
This also suggests the presence of semi-macroscopic defects that can produce large density fluctuations.
On the other hand, the local nature of the BBN strongly suggests that the BBN is generated by bulk pinning centers under the area of each sensor.

\subsubsection{Cross-correlation of the BBN and NBN; dependence on the array direction relative to the flow direction}

We also investigated the cross-correlation of the BBN and NBN in the direction perpendicular to the flow direction.
Figure~\ref{crossperp} shows the coherence $h_{ij}$ taken between two neighboring sites when the sensors were in the direction perpendicular to the flow direction.
Both for the BBN and the NBN, the cross-correlations are very small.
By comparing the BBN data in Figs.~\ref{crossperp}~(a)~and~(b) with those in Fig.~\ref{crosstotal}, it was found that the BBN showed larger correlations in the direction of the vortex flow than in the perpendicular direction at low frequencies below 100 Hz.
(At high frequencies, the noise level was very close to the background noise level.  Thus, we do not discuss the data at high frequencies.)
Thus, these data suggest that the coherence of the vortex lattice is better developed in the translational direction than in the transverse direction.
As will be discussed in the next section, this is consistent with the concept of channel-like flow characteristic of plastic flow.

This very large anisotropy was also found in the spatial correlation of the NBN.
As will be discussed later, however, we consider the meaning of the anisotropy in the spatial correlation to be very different between the BBN and the NBN.

\section{DISCUSSION}

\subsection{Origin of the BBN}

In the previous sections, we showed the various experimental features of the BBN data, some of which were those in the solid phase and the others were in the liquid phase.
We believe that even in the liquid phase the spatial correlation makes sense, for short-range correlations exist even in the liquid phase.  Thus, the discussion below is applicable both for the solid and liquid phases. 

To summarize the experimental features of the BBN, (1)  it is independent of current direction, (2) it shows the finger-print effect, and (3) it shows rather small spatial correlations even between neighboring sites (30 $\mu$m far).
All of these characteristics suggest that the BBN is generated by local-density fluctuations caused by bulk pinning centers which are present under the active areas of the Hall sensor.
We stress that this conclusion became convincing by the spatial correlation measurements.

The data presented in this paper showed that the fluctuations generated near site 3 show larger spatial correlations compared with other fluctuations.
As was discussed in the previous section, this large correlation is exceptional, and the general trend for the BBN is small spatial correlations.
On the other hand, the noise data from site 3, together with other correlation data, rather suggest that the correlation length of the BBN is something between 60-30 $\mu$m.

\subsection{Origin of the NBN}

In contrast to the BBN, the experimental features of the NBN may be summarized as follows. (1) The presence of the NBN itself is dependent on the samples.
(2) The time scale is the transit time of vortices.
(3) It is dependent on the current direction.
(4) Large translational correlations exist even between the farthest sites (90 $\mu$m).
These results suggest that the NBN is related to the semi-macroscopic transmission of the density fluctuation at some location in the sample.
Two candidates for the origin are possible.  One is a semi-macroscopic defect in the bulk.
Indeed, macroscopic linear defects were observed in this material\cite{Kob}.
The other candidate is surface.
The asymmetry observed under the sign reversal of the driving force strongly suggests that the surface explanation is more plausible.
In fact, in the presence of a surface barrier, the entry and exit of the vortices into/out of the sample was highly asymmetric\cite{Elli}.

\subsection{Plastic Flow}

Above, we discussed that the BBN had a bulk origin, whereas the NBN had an origin related to the presence of the surface barrier.
Thus, below, we concentrate on the BBN.

In previous studies of the noise generated by the vortices in superconductors (mostly the conduction noise), it was considered that the noise was generated when the vortices were depinned, or when the vortices underwent plastic flow\cite{Bat1,Danna,Safar,WJYeh}.
However, it was quite difficult to discriminate solely on these experiments which of the two mechanisms was dominant in each set of data.  In almost all cases of previous studies, information obtained by numerical simulations\cite{Dominiguez,Ryu,Olson} played an essential role to obtain conclusions.

In the present study, however, we measured the spatial correlation of the BBN directly for two different relative orientations (parallel and perpendicular to the flow direction).
Thus, it became possible to understand the dominant mechanism of the BBN generation solely based on the experimental data.

The magnitude of the spatial correlations of the BBN was different when the Hall probe array was placed parallel to the flow direction compared to perpendicular to the flow direction.
This clearly indicates that coherence of the vortex lattice developed better in the flow direction than in the perpendicular direction.
This is in contradiction to the elastic motion of the isotropic vortex bundle.
It is rather more consistent with channel-like motion considered in the concept of plastic flow.
Plastic flow was already observed in conventional superconductors\cite{TMatsuda}.
Our location-sensitive noise measurement also indicates that plastic flow occurs in high-$T_c$ superconductors.

Concerning on the anisotropy of the spatial correlations of the BBN, one may interpret the anisotropy simply in terms of the anisotropy of the elastic modulus of the vortex lattice. That is, the anisotropy may simply be due to the fact that the
long-wavelength compression modulus, $c_{11}$, of the vortex lattice is
much larger than the shear modulus, $c_{66}$, and may not be necessarily an
indication for plastic flow.
However, we do not think that the difference in the elastic modulus can explain the anisotropy in the noise data.
We measured the LDN in the steady state.  Thus, even if we have any differences between $c_{11}$ and $c_{66}$ (they may differ by 4$\sim$5 orders of magnitude in the present case, indeed), the fluctuation should have definite 
spatial correlations in the perpendicular direction as well as in the parallel direction, provided there is a finite 
$c_{66}$.  In other words, there will be no differences in the spatial correlation length of the fluctuation between these 
two directions. Since the anisotropy data were taken in the solid phase where  $c_{66}$ was finite, the noise data 
would behave similarly even in quantitatively, unless there was a plastic nature in the vortex motion.
Therefore, it is unlikely that the anisotropy in the noise data shown up in our experiments is interpreted by the difference in the elastic modulus.  Thus, the data strongly suggest that the plastic nature exists in the vortex motion.

The NBN was also found to have large anisotropy in the spatial correlations.
It should be noted, however, that the large anisotropy found in the spatial correlation of the NBN has a quite different meaning.
Since it already became clear that the NBN in the LDN is the manifestation of the translation of a semi-macroscopic density fluctuation created at the surface {\it etc.}, the anisotropy in the spatial correlation demonstrates that the fluctuation translates in one direction parallel to the vortex flow.
We do not think that the anisotropy in the spatial correlation of the NBN is related to the channel-like characteristics of the vortex motion.

\subsection{Correspondence to Theoretical Models}

Finally, let us discuss the relation between our noise data and existing theories of the dynamic states of vortices.

Through our noise study, the BBN was found to be the manifestation of the plastic motion of the vortices under the effects of random bulk pinning.
Since the BBN was largest just before the resistivity onset at all temperatures, it can be said that plastic flow was realized just before the resistivity onset.
This behavior is consistent with almost all the existing theories and simulations\cite{Vinokur,Balents,Doussal,Zimanyi,Dominiguez,Ryu,Olson}.
Thus, we could not restrict the theoretical possibilities based solely on the existing data of the LDN.
As was already mentioned in section I, the presence of the washboard conduction noise should force certain theories\cite{Balents,Zimanyi} to reconsider the dynamic state.
Without this other information, however, we could only state that our data are consistent with all the theories currently available.

It should be also noted that at all temperatures the noise is very small in the liquid state.
This is closely related to the presence of the FOT.
However, this is quite different from the dynamic melting/re-ordering transition proposed in ref.~\cite{Vinokur}.
In general, the noise power decreased rapidly with increasing driving force.
At present, we cannot separate the ``phase transition" from a rapid crossover from plastic flow to ordered flow.

According to a simulation\cite{Olson}, the crossover takes place via ``the  
smectic phase", where there are finite spatial correlations only  
in the flow direction.  Since the characteristic length scale of the
spatial correlations in the smectic phase is the vortex lattice
spacing, which is $\sim$ submicrons in the present situation, it is impossible to judge whether or not the smectic phase exists from our present apparatus.

\section{CONCLUSION}

Noise in the local vortex density was investigated in the mixed state of a high-$T_c$ superconductor, Bi$_2$Sr$_2$CaCu$_2$O$_y$, using 2DEG micro-Hall probe array.  We studied the cross-correlation function, together with the auto-correlation function, both parallel and perpendicular to the flow direction of vortices.
 The BBN exhibited the finger-print effect, and did not have large spatial correlations.  These observations suggest that the BBN is due to fluctuations of the local vortex density generated by the bulk pinning centers under the area of each probe.
On the other hand, the NBN, whose time scale was found to be the transit time of the vortices, had large translational correlations.  Various features of the NBN, including the very presence or absence of the NBN, depended on the samples.
Thus, we consider the NBN to have been generated by semi-macroscopic imperfections.
In relation to the dynamic phase diagram, large BBN was observed just when the vortices started moving.  The dependence of the spatial correlations on the direction of the array relative to the direction of the driving current suggested that plastic flow was realized when the large BBN was observed,
changing into coherent flow with further increasing driving current.
The gross features of our data agree well with some of the theoretically proposed dynamical phase diagrams of vortices in superconductors.

Although our noise measurements were on the local-density fluctuation, almost all other noise data have focused on the conduction noise, which mainly represents the velocity fluctuations averaged over the whole sample.
To get more detailed information from the noise experiments, it is valuable to compare these two different kind of fluctuation in the same sample.
This work is in progress, and the results will be published in a separate publication.

\section*{ACKNOWLEDGMENTS}
We thank F. Nori, C. J. Olson, T. Giamarchi, P. Le Doussal, D. Dominiguez, H. Matsukawa and J. R. Clem for fruitful discussions and D. G. Steel for a critical reading of the manuscript.
This work was, in part, supported by Grant-in-Aid for Scientific Research on Priority Area ``Vortex Electronics".
T. T, Y. T, and H. K thank the Japan Society for the Promotion of Science for financial support.


\newpage

\begin{table}
\twocolumn[\hsize\textwidth\columnwidth\hsize\csname@twocolumnfalse\endcsname
\begin{center}
\begin{tabular}{ccccccc}       \hline
sample \#   &  Dim. (mm$^3$) & $T_c$ (K)   &  $B_{FOT}$ (G)  &  $T$ dep. & location dep. & cross-correlation \\  \hline
\# 92r3 & 1.6 $\times$ 0.8 $\times$ 0.020 & 82 & 43 ($T$ = 70 {\rm K}) & $\circ$ & $\times$ & $\times$ \\
\# 72a1 & 0.4 $\times$ 0.2 $\times$ 0.020 & 87 & 26 ($T$ = 80 {\rm K}) & $\times$ & $\circ$ & $\times$ \\
\# 72c6 & 1.2 $\times$ 0.5 $\times$ 0.020 & 89 & 49 ($T$ = 80 {\rm K}) & $\times$ & $\times$ & $\circ$ \\  \hline
\end{tabular}
\end{center}
\caption{
Samples which were used in the measurements described in this paper.  $B_{FOT}$ is the magnetic field of the first-order phase transition of the vortex lattice in the equilibrium phase, and $T_c$ is the zero-resistance critical temperature of superconductivity.  $\circ$ and $\times$ represent that the measurement was performed and not performed, respectively.
}
\label{samples}
\vspace{10truemm}
]
\end{table}

\newpage

\begin{figure}
\vspace{20truemm}
\leavevmode
\caption{
Experimental setup for local-vortex-density-noise measurement and resistivity measurement.
}
\label{expsetup}
\end{figure}


\begin{figure}
\caption{
Schematic figures for (a) (array)$\parallel$(flow direction) configuration {\it v.s.} (b) (array)$\perp$(flow direction) configuration.
}
\label{expprlvsperp}
\end{figure}


\begin{figure}
\caption{
Dc resistivity and local magnetization at zero-current of Bi$_2$Sr$_2$CaCu$_2$O$_y$ crystals as a function of magnetic field under several fixed driving currents. (a) {\#}92r3 (b) {\#}72a1 (c) {\#}72c6.
}
\label{resist}
\end{figure}


\begin{figure}
\caption{
Density noise spectra of site 2 in sample {\#}72c6 as a function of current density at 31.8 Oe and at 65 K.
Current density was increased by 48 A/cm$^2$ in each spectrum.
The spectrum at zero current density is almost equal to the background noise spectrum (zero current density and zero magnetic field) in the present system.
}
\label{noisetotal}
\end{figure}


\begin{figure}
\caption{
Equi-noise contour (solid curves) {\it vs} equi-velocity contour (dashed curves) in the $H-j$ plane at 44 K (sample {\#}92r3).
The velocity of 7.5 $\times$ 10 cm/s corresponds to the resistivity of 10$^{-9}$ $\Omega$cm.
Inset: Equi-noise contour in the $H-j$ plane taken at various temperatures of sample {\#}92r3.
Magnetic fields correspond to the FOT are 151, 67 and 28 Oe for 50, 65 and 75 K, respectively, and the second-peak fields are 237, 224, and 197 Oe for 32, 38, and 44 K, respectively.
}
\label{BBNTdep}
\end{figure}


\begin{figure}
\caption{
Noise power spectral density of sample {\#}72a1 at a site at 80 K, taken at current densities with the same magnitude and the different directions.
The signs + and - in the figure represent that the current directions which are reverse to each other.
}
\label{BBNIinvert}
\end{figure}


\begin{figure}
\caption{
BBN power spectral density of sample  {\#}72c6 generated at each different site, when the sensor array was placed parallel to the flow direction.  Magnetic field, current density and temperature were 115 Oe, 160 A/cm$^2$, and 65 K, respectively.  In this figure, the background noise at zero-current density and zero magnetic field (shown in Fig. 4) were subtracted, for convenience.
Since the FOT took place at 134 Oe at this temperature, the data were taken in the solid state.
}
\label{BBNPSeach}
\end{figure}


\begin{figure}
\caption{
Coherence, $h$, of the BBN spectra in Fig. 7 taken between several different neighboring sites which are 30 $\mu$m apart when the sensor array is placed in the direction parallel to the flow direction.
Magnetic field, current density and temperature were 115 Oe, 160 A/cm$^2$, and 65 K, respectively. 
}
\label{crosstotal}
\end{figure}


\begin{figure}
\caption{
Coherence, $h$, of the BBN spectra in Fig. 7 taken between several different sites which are (a) 60 $\mu$m apart and (b) 90 $\mu$m apart when the sensor array is placed in the direction parallel to the flow direction.
Magnetic field, current density and temperature were 115 Oe, 160 A/cm$^2$, and 65 K, respectively. 
}
\label{crossfar}
\end{figure}


\begin{figure}
\caption{
NBN Power spectral density of sample {\#}72c6 generated at each different site.
Magnetic field, current density and temperature were 115 Oe, 640 A/cm$^2$, and 65 K, respectively. 
The background noise was sbtracted.
}
\label{NBNPSeach}
\end{figure}


\begin{figure}
\caption{
Coherence, $h$, of the NBN spectra in Fig. 10 taken between several different locations which are (a) 30 $\mu$m apart and (b) 60 $\mu$m apart and (c) 90 $\mu$m apart when the sensor array is placed in the direction parallel to the flow direction.
Magnetic field, current density and temperature were 115 Oe, 640 A/cm$^2$, and 65 K, respectively.
}
\label{NBNcrossfar}
\end{figure}


\begin{figure}
\caption{
Coherence, $h$, of the density noise taken between several different sites which are 30 $\mu$m apart when the sensors were put in the direction  perpendicular to the flow direction. (a) and (b) are for the BBN, and (c) and (d) are for the NBN.  (a) and (c) are between sites 1 and 2, and (b) and (d) are between sites 2 and 3.
Temperature and magnetic field were 65 K and 115 Oe, respectively, and current densities were 160 A/cm$^2$ for (a) and (b), and 640 A/cm$^2$ for (c) and (d), respectively. 
}
\label{crossperp}
\end{figure}



\begin{references}

\bibitem{Blatter}
G. Blatter, M. V. Feigel'man, V. B. Geshkenbein, 
A. I. Larkin, and V. M. Vinokur, 
Rev. Mod. Phys. {\bf 66}, 1125 (1994).

\bibitem{George}
 See for instance: G. W. Crabtree and D. R. Nelson, 
Phys. Today {\bf 50} (4), 38 (1997).

\bibitem{FOTBSC}
H. Pastoriza, M. F. Goffman, A. Arribere, and F. de la Cruz, 
Phys. Rev. Lett. {\bf 72}, 2951 (1994); 
E. Zeldov, D. Majer, M. Konczykowski, V. B. Geshkenbein, 
V. M. Vinokur, and H. Shtrikman, 
Nature (London) {\bf 375}, 373 (1995); 
T. Hanaguri, T. Tsuboi, A. Maeda, T. Nishizaki, N. Kobayashi, 
Y. Kotaka, J. Shimoyama, and K. Kishio
Physica C {\bf 256}, 111 (1996).

\bibitem{FOTYBC}
U. Welp, J. A. Fendrich, W. K. Kwok, G. W. Crabtree, and B. W. Veal,
Phys. Rev. Lett. {\bf 76}, 4809 (1996); 
A. Schilling, R. A. Fisher, N. E. Phillips, U. Welp, D. Dasgupta, 
W. K. Kwok, and G. W. Crabtree, 
Nature (London) {\bf 382}, 791 (1996); 

\bibitem{Tinkham}
M. Tinkham, Introduction to Superconductivity (McGraw-Hill, New York, 1975).

\bibitem{Vinokur}
A. E. Koshelev and V. M. Vinokur, Phys. Rev. Lett. {\bf 73}, 3580 (1994).

\bibitem{Balents}
L. Balents and M. P. A. Fisher, Phys. Rev. Lett. {\bf 75}, 4270 (1995); 
L. Balents, M. C. Marchetti, and L. Radzihovsky, Phys. Rev. B {\bf 57}, 7705 (1998).

\bibitem{Doussal}
T. Giamarchi and P. Le Doussal, Phys. Rev. Lett. {\bf 76}, 3409 (1996); Phys. Rev. B {\bf 55}, 6577 (1997); P. Le Doussal and T. Giamarchi, Phys. Rev. B {\bf 57}, 11356 (1998).


\bibitem{Balents2}
L. Balents, M. C. Marchetti and L. Radzihovsky, 
Phys. Rev. Lett. {\bf 78}, 751 (1997);
T. Giamarchi and  P. Le Doussal, 
Phys. Rev. Lett. {\bf 78}, 752 (1997).


\bibitem{Jensen}
H. J. Jensen, A. Brass, and A. J. Berlinsky, 
Phys. Rev. Lett. {\bf 60}, 1676 (1988).


\bibitem{Zimanyi}
K. Moon, R. T. Scalettar and G. T. Zim\'{a}nyi, 
Phys. Rev. Lett. {\bf 77}, 2778 (1996).


\bibitem{Dominiguez}
D. Dom\'{\i}nguez, N. Gr{\o}nbech-Jensen and A. R. Bishop, 
Phys. Rev. Lett. {\bf 78}, 2644 (1997). 


\bibitem{Ryu}
S. Ryu, M. Hellerqvist, S. Doniach, A. Kapitulnik, and D. Stroud, 
Phys. Rev. Lett. {\bf 77}, 5114 (1996).


\bibitem{Olson}
C. J. Olson, C. Reichhardt, and Franco Nori, 
Phys. Rev. Lett. {\bf 81}, 3757 (1998).


\bibitem{Gruner}
G. Gr\"{u}ner, Rev. Mod. Phys. 60, 1129 (1988); Density Waves in Solids (Addison-Wesley Longmans, Inc., Reading, 1994).

\bibitem{Mart}
P. Martinoli, O. Daldini, C. Leemann, and B. Van den Brandt, 
Phys. Rev. Lett. {\bf 36}, 382 (1976).

\bibitem{Troy}
A. M. Troyanovski, J. Aarts, and P. H. Kes, Nature (London) {\bf 399}, 665 (1999).



\bibitem{Togawa}
Y. Togawa, R. Abiru, K. Iwaya, H. Kitano, and A. Maeda, 
Phys. Rev. Lett. {\bf 85}, 3716 (2000). 

\bibitem{Shobo1}
S. Battacharya and M. J. Higgins, 
Phys. Rev. Lett. {\bf 70}, 2617 (1993); Phys. Rev. B {\bf 52}, 64 (1995).


\bibitem{Henderson}
W. Henderson, E. Y. Andrei, M. J. Higgins, and S. Bhattacharya, 
Phys. Rev. Lett. {\bf 77}, 2077 (1996); 
W. Henderson, E. Y. Andrei, and M. J. Higgins, 
{\it ibid}. {\bf 81}, 2352 (1998);
Z. L. Xiao, E. Y. Andrei, and M. J. Higgins, 
{\it ibid}. {\bf 83}, 1644 (1999); 
Z. L. Xiao, E. Y. Andrei, P. Shuk and M. Greenblatt, 
{\it ibid}. {\bf 85}, 3265 (2000); 
{\it ibid}. {\bf 86}, 2431 (2001).

\bibitem{Yaron}
U. Yaron, P. L. Gammel, D. A. Huse, R. N. Kleiman, C. S. Oglesby, 
E. Bucher, B. Batlogg, D. J. Bishop, K. Mortensen, and K. N. Clausen, 
Nature (London) {\bf 376}, 753 (1995).


\bibitem{Pardo}
F. Pardo, F. de la Cruz, P. L. Gammel, E. Bucher, and D. J. Bishop,
Nature (London) {\bf 396}, 348 (1998).


\bibitem{Fendrich}
J. A. Fendrich, U. Welp, W. K. Kwok, A. E. Koshelev, 
G. W. Crabtree, and B. W. Veal, 
Phys. Rev. Lett. {\bf 77}, 2073 (1996).

\bibitem{Tsuboi1}
T. Tsuboi, T. Hanaguri, and A. Maeda, Phys. Rev. B {\bf 55}, R8709 (1997).

\bibitem{TMatsuda}
T. Matsuda, K. Harada, H. Kasai, O. Kamimura, and A. Tonomura,  
Science {\bf 271}, 1393 (1996). 


\bibitem{Tsuboi2}
T. Tsuboi, T. Hanaguri, and A. Maeda, Phys. Rev. Lett. {\bf 80}, 4550 (1998).

\bibitem{Tsuboishort}
T. Tsuboi, T. Hanaguri, A. Maeda, R. Abiru, K. Iwaya, and H. Kitano,
Physica {\bf B 284-288}, 843 (2000).

\bibitem{Maeda}
A. Maeda , T. Tsuboi, T. Hanaguri, Y. Togawa, R. Abiru, 
Y. Tsuchiya and K. Iwaya,
J. Low Temp. Phys. {\bf 117}, 1329 (1999).

\bibitem{Clem}
For a review of the noise study in conventional superconductors, see for example, J. R. Clem, Phys. Reports {\bf 75}, 1 (1981).

\bibitem{Bat1}
A. C. Marley, M. J. Higgins, and S. Battacharya, 
Phys. Rev. Lett. {\bf 74}, 3029 (1995).

\bibitem{Bat2}
R. D. Merithew, M. W. Rabin, M. B. Weissman, 
M. J. Higgins and S. Bhattacharya, 
Phys. Rev. Lett. {\bf 77}, 3197 (1996).

\bibitem{Danna}
G. D'Anna, P. L. Gammel, H. Safar, G. B. Alers, D. J. Bishop, 
J. Giapintzakis and D. M. Ginsberg, 
Phys. Rev. Lett. {\bf 75}, 3521 (1995).

\bibitem{Safar}
H. Safar, P. L. Gammel, D. A. Huse, G. B. Alers, D. J. Bishop, 
W. C. Lee, J. Giapintzakis, and D. M. Ginsberg, 
Phys. Rev. B {\bf 52}, 6211 (1995).

\bibitem{Paltier}
Y. Paltiel, E. Zeldov, Y. N. Myasoedov, H. Shtrikman, S. Bhattacharya, 
M. J. Higgins, Z. L. Xiao, E. Y. Andrei, P. L. Gammel, and D. J. Bishop,  
Nature (London) {\bf 403}, 398 (2000); 
Y. Paltiel, E. Zeldov, Y. Myasoedov, M. L. Rappaport, G. Jung, S. Bhattacharya, M. J. Higgins, Z. L. Xiao, E. Y. Andrei, P. L. Gammel, and D. J. Bishop, 
Phys. Rev. Lett. {\bf 85}, 3712 (2000).



\bibitem{Gordiev}
S. N. Gordeev, P. A. J. d. Groot, M. Oussena, A. V. Volkozub, 
S. Pinfold, R. Langan, R. Gagnon, and L. Taillefer, 
Nature (London) {\bf 385}, 324 (2000); 
S. N. Gordeev, A. P. Rassau, P. A. J. d. Groot, 
R. Gagnon, and L. Taillefer, 
Phys. Rev. B {\bf 58}, 527 (1998).


\bibitem{WJYeh}
W. J. Yeh, L. K. Yu, M. Yang, L. W. Song, and Y. H. Kao , 
Physica C {\bf 195}, 367 (1992); 
W. J. Yeh, and Y. H. Kao ,
Phys. Rev. B {\bf 44}, 360 (1991).

\bibitem{Kob}
M. R. Koblischka, R. J. Wijngaarden, D. G. de Groot, R. Griessen, 
A. A. Menovsky and T. W. Li, 
Physica {\bf C249}, 339 (1995).


\bibitem{Elli}
E. Zeldov, A. I. Larkin, V. B. Geshkenbein, M. Konczykowski, D. Majer, 
B. Khaykovich, V. M. Vinokur, and H. Shtrikman, 
Phys. Rev. Lett. {\bf 73}, 1428 (1994); 
D. Majer, E. Zeldov, M. Konczykowski, V. B. Geshkenbein, A. I. Larkin, 
L. Burlachkov, V. M. Vinokur, and N. Chikumoto, 
Physica {\bf C 235-240}, 2761 (1994).



\end{references}
\end{document}